# WiNNbeta: Batch and drift correction method by white noise normalization for metabolomic studies.

Olga Demler*, Franco Giulianini*, Yanyan Liu, Malte Londschien, Anja Sjöström, Tanmay Tanna, Heike Luttmann-Gibson, Antoine Jeanrenaud

## Abstract

Recent advances in chemical analytics techniques such as Liquid Chromatography-Mass Spectrometry (LC-MS), Gas Chromatography (GC)-MS and others allow measurements of low concentrations of thousands of metabolites in blood and other tissues. However, such high-throughput measurements are prone to batch effects and drifts, resulting in high coefficients of variation (CVs) for a significant fraction of measured metabolites. As much as 40% of metabolites may be discarded because of high CVs. We assume that in a well-designed metabolomics experiment relative intensity versus "run order" (i.e. the order in which samples are processed) should be independent and identically distributed, i.e. follow properties of white noise. We developed a method called batch and drift correction method by White Noise Normalization (WiNNbeta) to correct individual metabolites for batch effects and drifts. This method tests for white noise properties to identify metabolites in need of correction and corrects them by using fine-tuned splines. To test the method's performance we applied WiNNbeta to LC-MS data from our metabolomic studies and computed CVs before and after WiNNbeta correction in quality control samples (human pooled plasma). In our tests, WiNNbeta increased the percent of metabolites with CVs less than .2 from 33% to 40%. Three metabolites were also measured by CLIA-lab using traditional assays. Correlations between CLIA-lab and LC-MS measurements corrected by WiNNbeta improved for all three metabolites. WiNNbeta can be applied to a wide range of omic measurements, it does not rely on quality control samples and filters out flat batch effects and drifts conservatively when appropriate.

## Introduction

Technological advances in chemical-analytical processing and instrumentation in the last 20 years resulted in greatly increased sensitivity of GC-MS and LC-MS instruments, which are now able to detect small molecules concentrations as low as < 0.01% of the metabolites with the highest intensity[1]. Metabolomics is a study of small molecules usually with molecular weight <1,500 Da that are present in a variety of tissues such as blood, urine, breast milk. Applications of metabolomics include monitoring of clinical trials, drug and biomarker discovery, among other applications[2]. Improved ability to detect low concentrations of small molecules in a variety of tissues, faster processing times coupled with cost efficient metabolite extraction techniques revolutionized metabolomics, which is now a fast-growing field. Large metabolomic studies often involve analysis of thousands of tissue samples that can take multiple days to weeks to process. In such large-scale studies, metabolomic measurements are prone to batch effects and instrumental drifts (Figure 1)[1,3,4]. In a typical LC-MS experiment, pre-processed human plasma samples are placed in the 96- or 384-well plates, which are then processed, one at a time, according to a specific run order. There may be many thousands of experimental samples in a single study, so several plates are processed sequentially in an analytical chemistry lab. Technical variation in experimental conditions and instrument drifts introduce batch effects (flat shifts in plate-by-plate measurements and drifts - within-plate monotone shifts in relative intensity)[1,4]. The presence of batch effects and drifts can be clearly seen by simply plotting metabolite relative intensity versus instrument run order. In Figure 1, for example, we plot intensities vs. run order for three different metabolites measured in three different studies: two LC-MS experiments with N=400 and N=3800 and in the Nuclear Magnetic Resonance (NMR) study (N=1600).

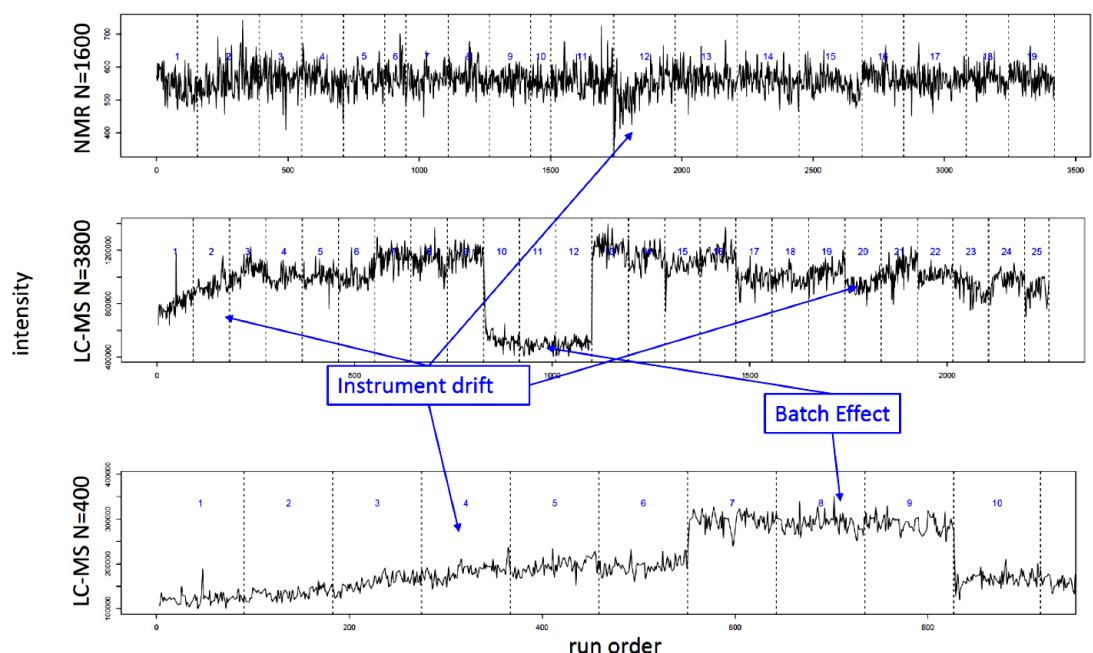

**Figure 1.** Relative intensities of one of the metabolites from three different metabolomic experiments plotted versus run order. 1,600 experimental samples processed using NMR technique. 3,800 plasma samples were placed on 36 plates and processed in an LC-MS experiment (first 25 are shown), 400 plasma samples were placed on 11 plates.

Figure 1 is a typical example of drifts and batch effects present in a real-life large-scale metabolomic data. Watrous et. al. discuss potential causes of batch effects and drifts in an LC-MS experiment[5] (for example column replacements or cleaning or any other adjustments usually performed by analytical chemist between plates). Plates represent natural points or "discontinuity" in an LC-MS experiment ("batch effects") while instrument drifts may be caused by metabolite degradation due to their chemical instability (downward drift) or contamination of the column (upward drift)[1,4].

Batch effects and drifts are artefacts of chemical extraction processes that can completely overwhelm the true signal. There are two ways to deal with batch effects and drifts. The first option is to discard from the study metabolites with large unwanted variability. However, this can lead to removal of up to 85% of metabolites. The second option is to correct the metabolomic measurements for batch and drift effects. There are a number of signal-correction methods (also called normalization methods) currently in use. The simplest ones are batch-correction methods by mean centering, median scaling. A commonly used normalization method in omics experiments is "ComBat" (Johnson et.al.[6]), which uses empirical Bayes hierarchical models to estimate batch correction. Reisetter et. al. introduced mixture model normalization method for signal correction[6]. Sysi-Aho et. al. developed a normalization method called NOMIS[7], which assumes that a metabolite measurement follows a normal distribution, where the mean is a composite of the true mean plus drift plus error. Error absorbs well-to-well variation. NOMIS assumes that drifts can be removed by a linear combination of internal standards. Karpiewitch et. al. introduced the EigenMS method, which uses singular value decomposition of analysis of variance (ANOVA) residuals to estimate and filter out drifts[8,9]. Dunn et. al. fit low-order nonlinear locally estimated smoothing splines (LOESS) to the quality control data with respect to the order of injection (QC-RLSC)[1]. Liu et. al. used wavelets to estimate drift correction[10]. Fan et. al. used random forest to remove unwanted variation[11]. However, none of these methods have three properties that are necessary for a good signal correction method. A good batch and drift correction method should have the following qualities:

*1) Correction should be performed only when needed:* as demonstrated in Watrous et. al.[4] not all metabolites require batch and/or drift correction.

*2) Correction should be metabolite-specific and should not affect other metabolites:* as demonstrated in Figure 1 and in Watrous et. al.[4] and Dunn et.al.[1] a number of different drift profiles can be observed in metabolomics data, which may result in distortion of the true signal.

*3) Correction should be conservative – it should not over-correct.*
In order to preserve true associations, it is important not to distort true relative intensities of metabolites by overcorrecting.

Here we introduce a new batch and drift correction method by white noise normalization, abbreviated as WiNNbeta. The method is based on the assumption, that given random placement of samples in a plate, the signal represented by the sequence of measurements should be serially uncorrelated, i.e. have "white noise" like characteristics. The presence of autocorrelations in the signal is assumed to indicate the presence of drifts and batch effects, which our method can detect and correct. WiNNbeta tries to strike a balance between drift correction and preservation of the original signal and since it does not rely on quality control (QC) samples, these can be used to evaluate how the method performs by comparing before- and after- correction CVs.

In the next sections we will provide a general description of the WiNNbeta algorithm, apply it to real and synthetic data and compare its performance with other normalization or drift correction techniques. WiNNbeta is implemented in R[12] and is available as a package.

**Methods**

*WiNNbeta: General Description*

LCMS experiments should be designed in such a way as to reduce or eliminate the effects of those factors which introduce drifts and biases in the data. One of such factors is sample placement in the plates, which should be completely randomized, so that sample characteristics are independent of the measurement run order. The intensity value of any given metabolite versus its run order should then, ideally, behave as an independent and identically distributed (idd) random variable, i.e. as white noise.

This idea that measurements from a good metabolomic experiment should have white noise characteristics is at the basis of our drift correction method. We use white noise testing to guide us in identifying metabolites that need drift-correction and in optimizing this correction. Another important postulate is that drift correction method should be very conservative (in order to preserve true associations) and should not overfit.

To summarize, WiNNbeta is predicated on the following 3 assumptions:

(1) Pre-processed experimental medium (plasma, urine, etc.) are injected <u>in random order</u> into the wells. Wells are processed by the instrument one by one. Therefore, this sequence of measurements should resemble white noise (WN). By random order we mean that the order in which samples are run through the instrument is not associated with any sample characteristics. For example, if we study downstream metabolites of an experimental drug in a randomized controlled trial with two treatment groups: active treatment group and a placebo group, we do not want to run through the instrument samples which share the same characteristics first and then all samples that all share another characteristics. This is not a random order. Instead we would randomly place samples so that there is no association of run order with any patient characteristic.

(2) Equipment drifts and/or changes in experimental conditions introduce trends in the sequence of measurements, which results in the loss of WN quality.

(3) Any tuning of the instrument can result in batch effects and discontinuities in measurements. Mapping of samples to their run order must be available. How run order is assigned to batches may be available. Therefore, the WiNNbeta algorithm will work if no batches are present (i.e., all samples belong to the same batch) but mapping of samples to run order is a necessary condition for WiNNbeta algorithm to work.

*WiNNbeta: Workflow*

WiNNbeta takes as an input the sequence of metabolite measurements $\{m_i\}$ and its workflow consists of 2 sequential distinct phases:

1. Removal of batch effects
2. Detrending

Phase 1: Removal of batch effects
To correct for batch[a] effects, WiNNbeta applies *variance normalization* and/or *residualization* to the input sequence $\{m_i\}$. Definition of a batch is provided below.

*Variance normalization*: the input sequence $\{m_i\}$ is tested for homogeneity of variance across plates (Levine or Fligner-Killean test). If the test fails, the metabolite is then normalized by the standard deviation of each plate: $\{m_{Ni}\} = \{m_i\}/sd(plate)$.

*Residualization*: in this step ANOVA (with "plate" as grouping variable) is performed on the $\{m_i\}$ (or $\{m_{Ni}\}$ if variance-normalization was applied) sequence. If the test fails, the metabolite is then residualized by plate according to the following regression model: $\{m_i\} = \beta_0 + \beta_1 * plate + \varepsilon_i$.
If the metabolite is residualized, then the $\{m_i\}$ (or $\{m_{Ni}\}$) is replaced by the corresponding $\{\varepsilon_i\}$ (or $\{\varepsilon_{Ni}\}$) sequence.

At the end of this phase, the sequence of measurements can, therefore, be in one of the following states:

(i) No variance normalized   + no residualized      $\{m_i\}$
(ii) Variance normalized     + no residualized      $\{m_{Ni}\}$
(iii) No variance normalized + residualized         $\{\varepsilon_i\}$
(iv) Variance normalized     + residualized         $\{\varepsilon_{Ni}\}$

Phase 2: Detrending
WiNNbeta operates on the assumption that if, after batch correction, the sequence of measurements fails WN testing, it is most likely because of a trend in the data. WiNNbeta tests for WN each plate. If $\{y_i\}$ is one of the sequences (i) - (iv) above, entering detrending after failing the WN test for a given plate, then WiNNbeta assumes: $\{y_i\} = \{y_{detrend_i}\} + \{trend_i\}$,
where $\{y_{detrend_i}\}$ would be the final detrended sequence.

To determine $\{trend_i\}$, WiNNbeta applies a plate by plate unpenalized spline regression fit on $\{y_i\}$, with smoothing parameters optimally tuned to minimize autocorrelations in the signal (see Appendix Figure A4 and B1 for more details). Once $\{trend_i\}$ is computed, then $\{y_{detrend_i}\} = \{y_i\} - \{trend_i\}$.

Phase 3: Second removal of batch effects
WiNNbeta repeats Phase 1 bacth correction steps to remove residual batch effects that may be masked by drifts.

*Definition of a batch:* in this context, plates or batches are crudely defined as any natural grouping of the experimental sequence. It can be a plate or a date/hour etc. when samples were processed. Between plates / batches analytical chemist may perform cleaning of the instrument or make some adjustment, that may potentially result in abrupt shifts in metabolite intensity or batch effects. Therefore, batches define natural points of discontinuity within the data.

*White Noise Test*:  Test for white noise uses Box-Ljung test of presence of autocorrelations.

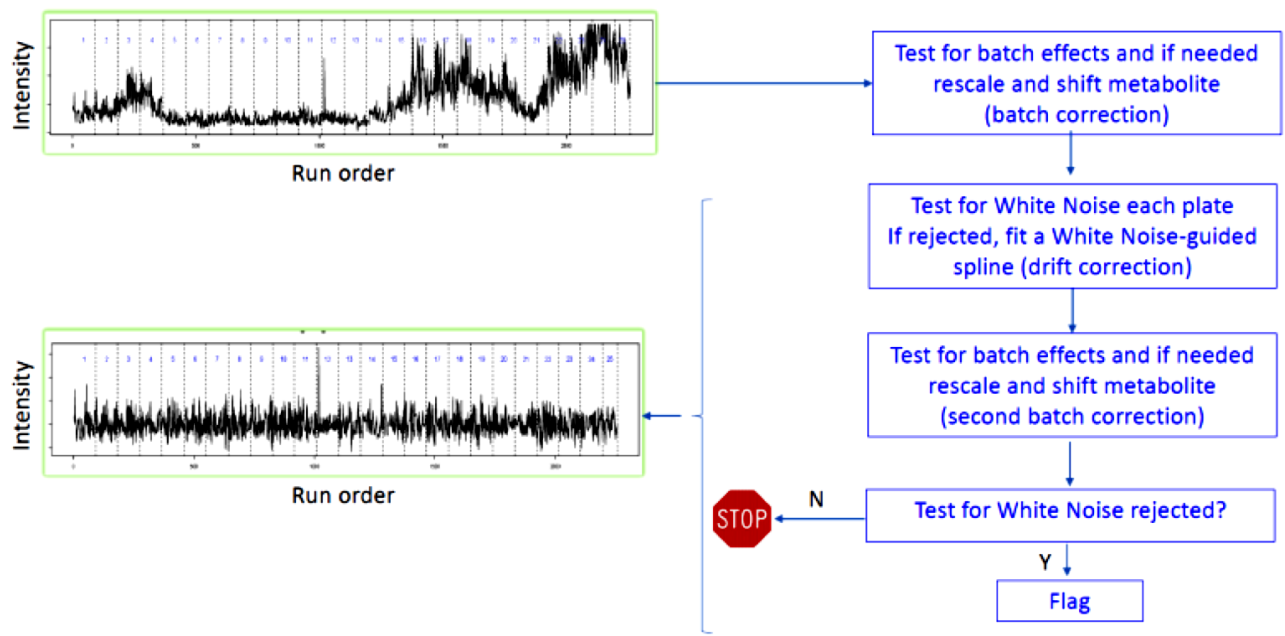

**Figure 2.** Outline of WiNNbeta batch and drift correction algorithm.

## Results

*Performance of WiNNbeta in LCMS and NMR data.*

We applied WiNNbeta to data from an LCMS experiment consisting of 1,581 metabolites measured in 1541 human plasma samples placed in 25 96-well plates. Of the 1,578 metabolites 93% required normalization (failed test of the equality of variance across plates), 97% failed the test of equality of means across plates and were also residualized by plate. After these two corrections were applied, 31% of metabolites failed the white noise test and further spline-based detrending was applied (the rest 69% passed the first white noise test and no further correction was done on them). Of those that were detrended, 26% passed the second white noise test. The before- and after-correction profiles for one of the metabolites is illustrated in Figure 3 (left column). Figure 3 shows also profile of a metabolite before and after correction using data from one of our NMR study (right column).

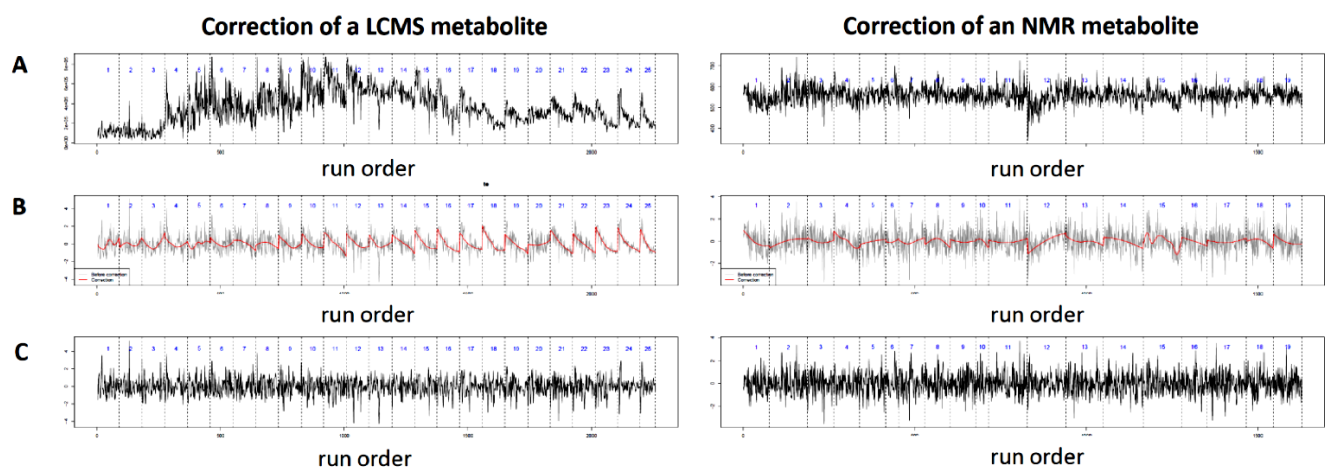

**Figure 3**. WiNNbeta-correction of NMR and LC-MS metabolomics data. (A): original signal; (B): n grey is the signal normalized by plate standard deviation and residualized by plate, in red is the spline-based trend found; (C) the corrected signal.

*Performance of WiNNbeta in simulated data*

We simulated "white noise" data by creating a sequence of independent and identically distributed (idd) normal random variables. The idd sequence was then split into 10 plates and various distortions were added to it.

Figure 4A shows the before- and after- correction of an iid signal plus a sinusoidal distortion while Figure 4B shows the same before- and after- correction profile of an idd + a piecewise-linear drift (more examples are provided in Appendix A1 a-g).

April 11, 2024

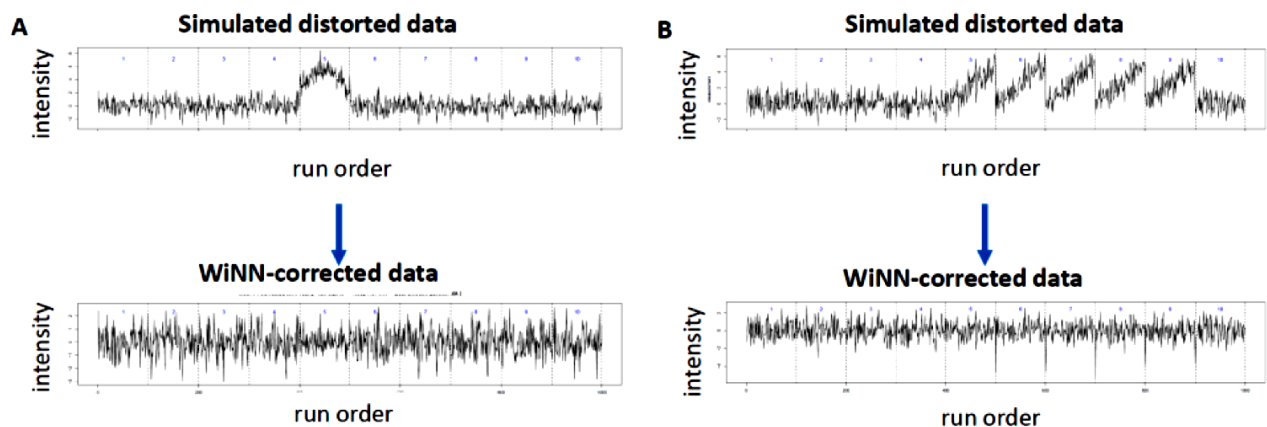

**Figure 4. WiNNbeta-correction of simulated white noise data (iid random variables) distorted by adding A. non-linear drift to one of the plates, B. piecewise-linear drifts.**

We calculated an error as a difference between WiNNbeta-corrected data and true measurements. The boxplot of the error in the series of simulations is presented in Figure 5.

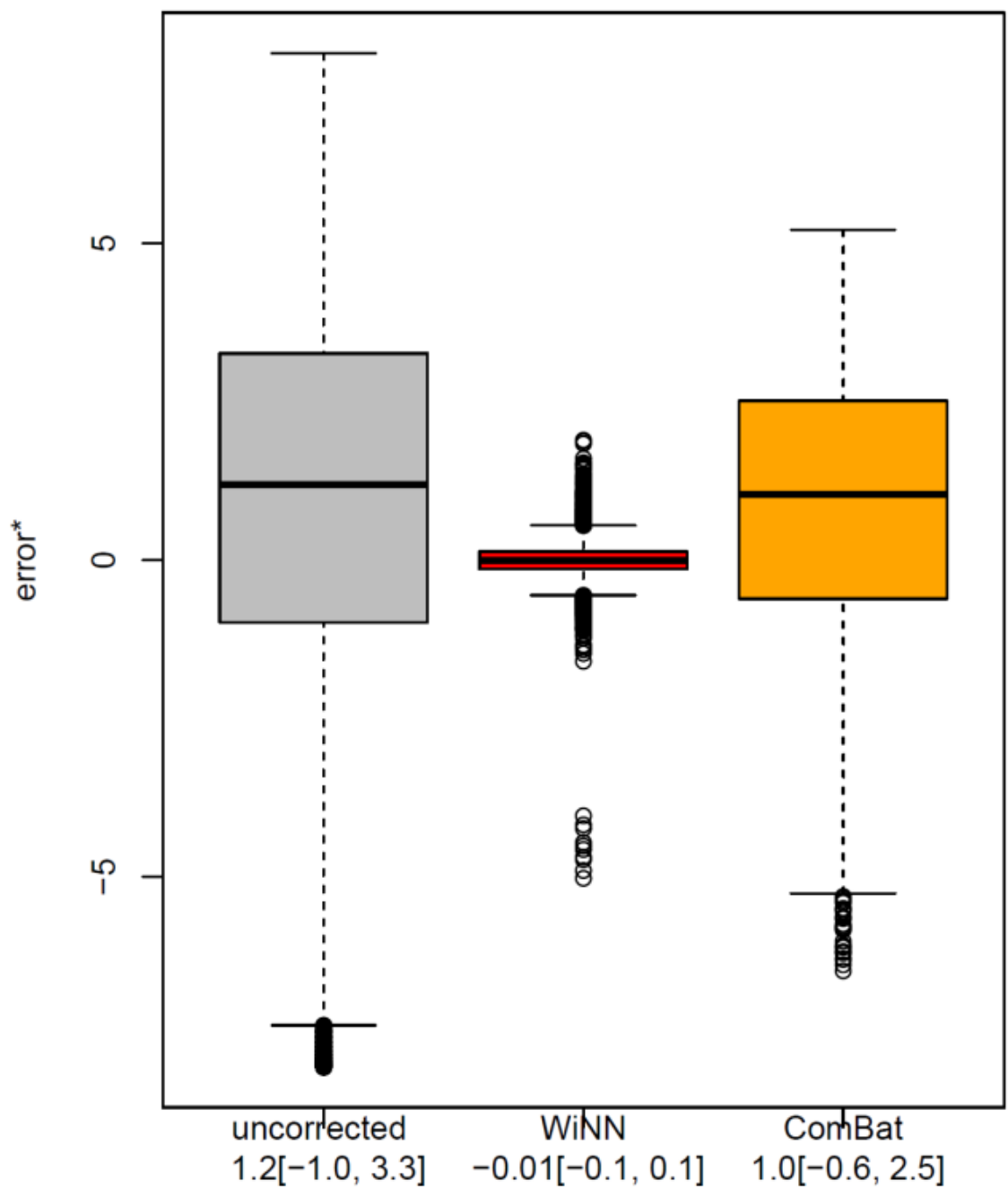

**Figure 5. Boxplots of error before and after WiNNbeta and Combat corrections in simulated data.**
*error is calculated as the difference between true measurement and distorted uncorrected (left) or WiNNbeta-corrected (middle) or ComBat-corrected (right) measurements. Under the boxplot labels are medians and interquartile ranges of the errors. Simulated data was generated by distorting the true measurement (N(0,1)) by adding various drifts (please see Appendix).

*Improvement of CVs*

To obtain an objective measure of improvement, we evaluated the performance of WiNNbeta in Quality Control (QC) samples. QC samples, such as pooled plasma, are usually added to each plate/batch to monitor performance of the equipment and are used to calculate quality control metrics[13]. The most commonly used metric for quality measurements is the coefficient of variation (CV), also called relative standard deviation (RSD). In our data QC samples consist of pooled plasma samples. Because these QC samples use identical blood pool, each metabolite should have minimal variability across them. We applied correction to pooled plasma data and compared CVs before and after the correction. In Figure 5 shows cumulative distribution of CV before and after WiNNbeta correction. FDA guidelines for CLIA labs state that the CV should be less than 20% for a good measurement[14]. In our LCMS data 523 metabolites had met the CV<20% criteria before correction; after the WiNNbeta correction, 633 met this criterion. Details of how to use Quality Control samples to calculate CVs before and after correction are presented in Appendix B1, implemented in R code which is available as part of the WiNNbeta package. Cumulative distribution curves before and after WiNNbeta correction are plotted in Figure 5.

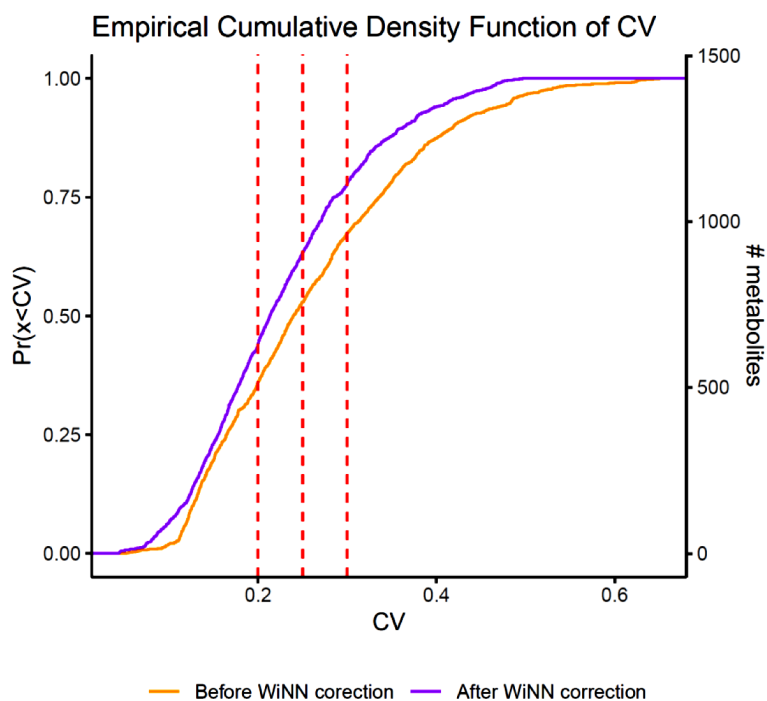

**Figure 6**. Cumulative distribution curves of CVs before and after WiNNbeta correction. We used data with N=1540 experimental samples with 1432 metabolites measured in each sample by LC-MS. Orange curve: before WiNNbeta correction, purple: after WiNNbeta correction.

*Improvement of correlation of metabolites measured in two independent experiments.*

In one of our experiments, three fatty acids (eicosapentaenoic acid (EPA), docosahexaenoic acid (DHA) and arachidonic acid (AA)) were measured twice – once using nontargeted LCMS assay and independently in a different laboratory with standard assaying technique - targeted phospholipid LC-MS2 assay (Supplementary Material B6). In Table 1 we present correlations between the nontargeted assay and the targeted assay before and after WiNNbeta correction. All required normalization and residualization and two required detrending.

April 11, 2024

**Table 1**. Spearman correlation of true measurements with original (uncorrected), WiNNbeta-corrected and ComBat-corrected.

|          | EPA | DHA | AA  | met1 | met2 | met3 | met4 | met5 | met6 | met7 | met8 | met9 | met10 | met11 |
|---------:|-----|-----|-----|------|------|------|------|------|------|------|------|------|-------|-------|
| original | .71 | .47 | .29 | .55  | .55  | .60  | .29  | .47  | .26  | .69  | .88  | .49  | .40   | .35   |
| WiNN     | .76 | **.53** | .32 | **.98** | **.98** | **.98** | **.95** | **.99** | **.96** | **.98** | **.99** | **1.00** | **1.00** | **.99** |
| ComBat   | **.76** | .53 | **.33** | .84  | .85  | .78  | .52  | .99  | .52  | .81  | 1.00 | 1.00 | .99   | .99   |

\* LCMS = nontargeted LC-MS metabolomics assay measuring fatty acids, oxylipins and bioactive lipid metabolite, prone to batch effects and drifts.
\*\* no subscript stands for targeted plasma phospholipid LC-MS2 assaying technique.

*Comparison of WiNNbeta to other drift-correction methods.*

We applied various signal correction methods to our LC-MS data. One way to check the quality of drift-correction is to see whether there is any clustering by plate before and after batch- and drift- correction. We checked for presence of clustering using "t-distributed stochastic neighbor embedding" (t-SNE)[15,16]. We plotted t-SNE components before and after different batch- and drift- correction methods in Figure 6. Each point in the Figure 6 relies on metabolite measurements obtained from one well. We colored each well by its plate number. In plots of original data in Figure 6 with raw data we can see clusters of wells that coincide with their plate numbers, this clustering is attenuated by NOMIS (we can still see a cluster of plate 5 in the top scatterplot) and QC-RLC. Clustering is much worsened by a running-median method. No clustering remains after WiNNbeta is applied.

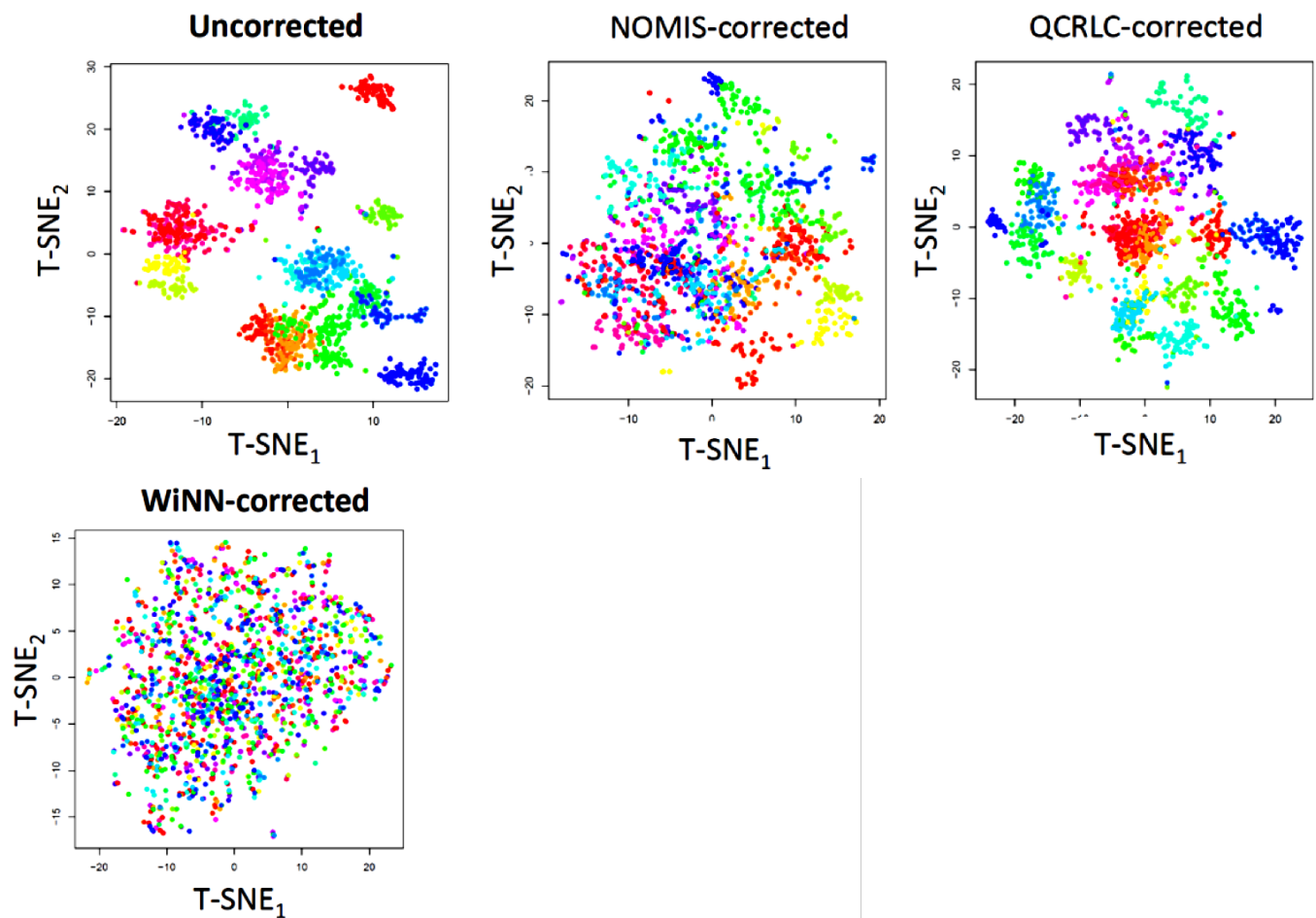

**Figure 7.**

It is important to note the difference between batch-correction methods and drift correction methods. Figure 7 illustrates that ComBat is a batch-correction method and has been presented as such in the original publication[17]. To illustrate that ComBat cannot correct for drifts, we used a sine function to distort simulated iid data and then added batch effects at each plate on top of it. We applied ComBat correction and WiNNbeta correction to this data and present intensity versus run order before (Figure 7A) and after correction (Figure 7C). CombBat levels out means across plates (batches) but it was not designed to pick up drifts within batches (plates) (Figure 7C).

April 11, 2024

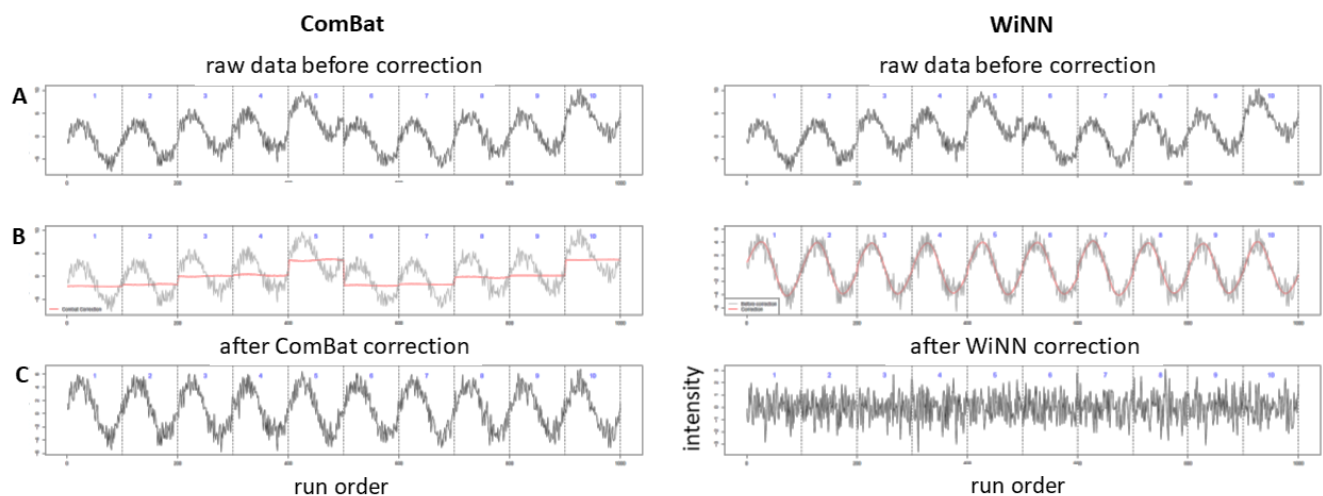

**Figure 8. Comparison of ComBat and WiNNbeta.** Rows: A – distorted data by adding sinusoidal drift + batch effect to true signal data (black); B – data (in black) with correction in red overlayed over it; C – data after correction. ComBat successfully removed batch effects but did not correct the drifts.

**Conclusions**
In this paper we developed a novel batch- and drift- correction method called WiNNbeta. It relies on properties of white noise to guide the decision if the correction is necessary and to estimate tuning parameters of the correction. This method performs well for a variety of analytic techniques used to extract metabolites. We compared correlations before and after WiNNbeta correction of three metabolites that were measured with LCMS and CLIA-lab-quality standard assaying technique and observed improvement in their correlation after WiNNbeta correction of LCMS metabolites.
Limitations of the WiNNbeta method are the following: WiNNbeta relies on knowledge of run order which is an order in which samples (blood, other tissue) are processed by the metabolomics instrument, this order can also be informed by knowledge of a sequence of observations that might be associated with spurious drifts and autocorrelations. WiNNbeta does not handle missing observations, which should be removed or imputed beforehand. WiNNbeta requires a minimum number of observations per batch (about 20). WiNNbeta is sensitive to outliers, which we recommend should be trimmed beforehand.
Strengths of WiNNbeta include:
*WiNNbeta does not use quality control samples.* QC samples in metabolomics studies are usually used to filter out metabolites with low quality of measurements (high CV); to monitor performance of the metabolomics instrument during sample processing and to correct for drifts. Estimation of CV requires fewer QC samples (i.e. 3 per plate etc.) while drift correction based on QC samples heavily depends on quantity of QC samples. Drift correction methods such as NOMIS, QC-RLC etc. that rely on QC samples may require adding as many as 20% of QC samples to the metabolomics experiment. WiNNbeta does not rely on QC samples which frees up the resources.
*WiNNbeta can be used during processing of samples.* WiNNbeta can detect presence of drifts and systematic batch effects and can be used to alert the technician that adjustments are necessary.

*WiNNbeta performs only necessary adjustments.* WiNNbeta tests each metabolite whether is deviates from white noise and correction is applied only when it is necessary.

WiNNbeta uses splines with estimated degrees of freedom that optimally restore the signal.

*WiNNbeta was validated in samples from an independent experiment* – WiNNbeta improved correlation of LC-MS EPA, DHA and AA with their concentrations measured in an independent experiment with standard assay techniques. *WiNNbeta reduced CVs, increasing the # of metabolites with CV≤20%. WiNNbeta can be applied to other omics data.*

**Appendix**

**Figure A1 Filtering out drifts and batch effects by WiNNbeta.**

A. Added drifts to some plates

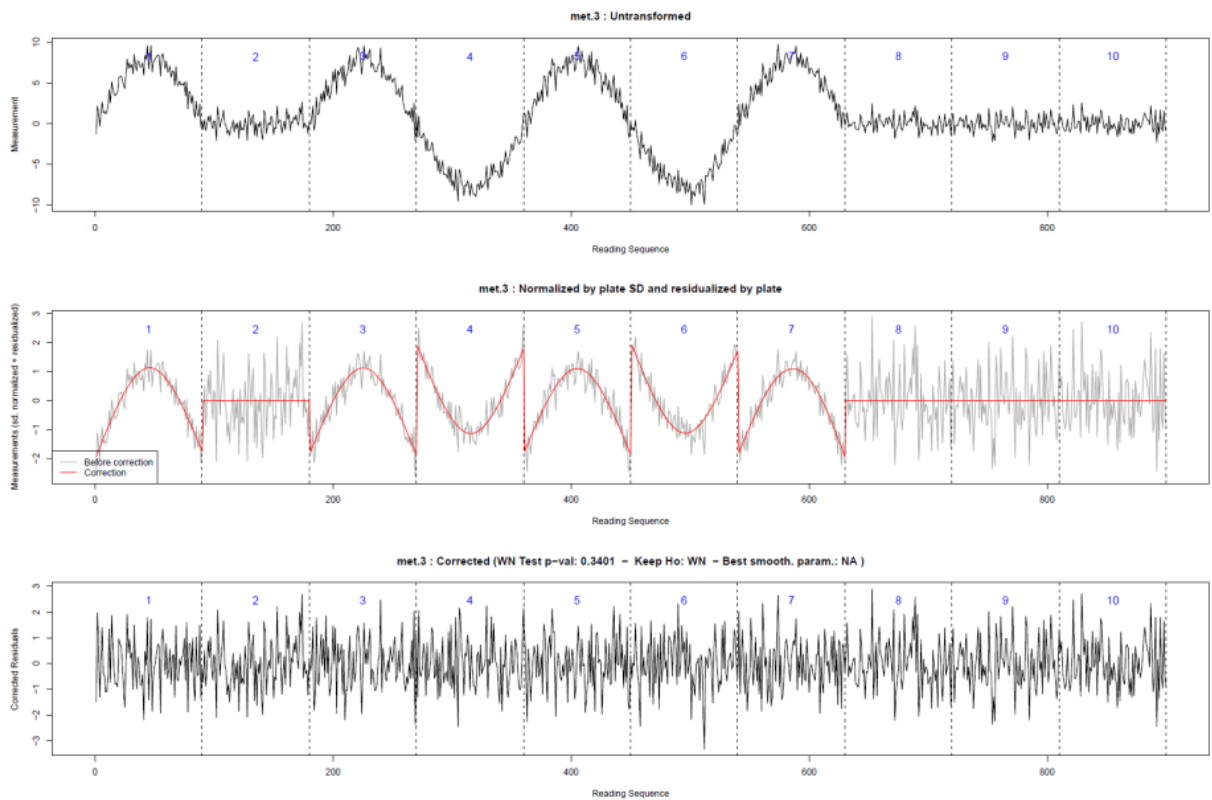

**B. Added piecewise-linear drifts to some plates**

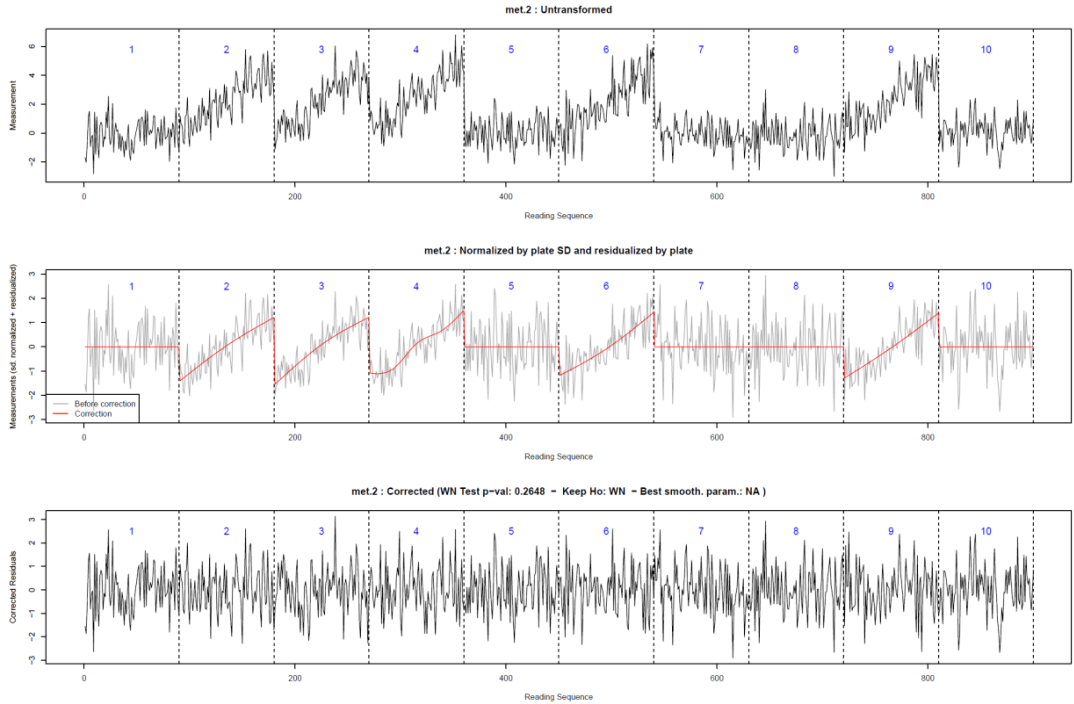

**C. Added non-linear drifts to each plate**

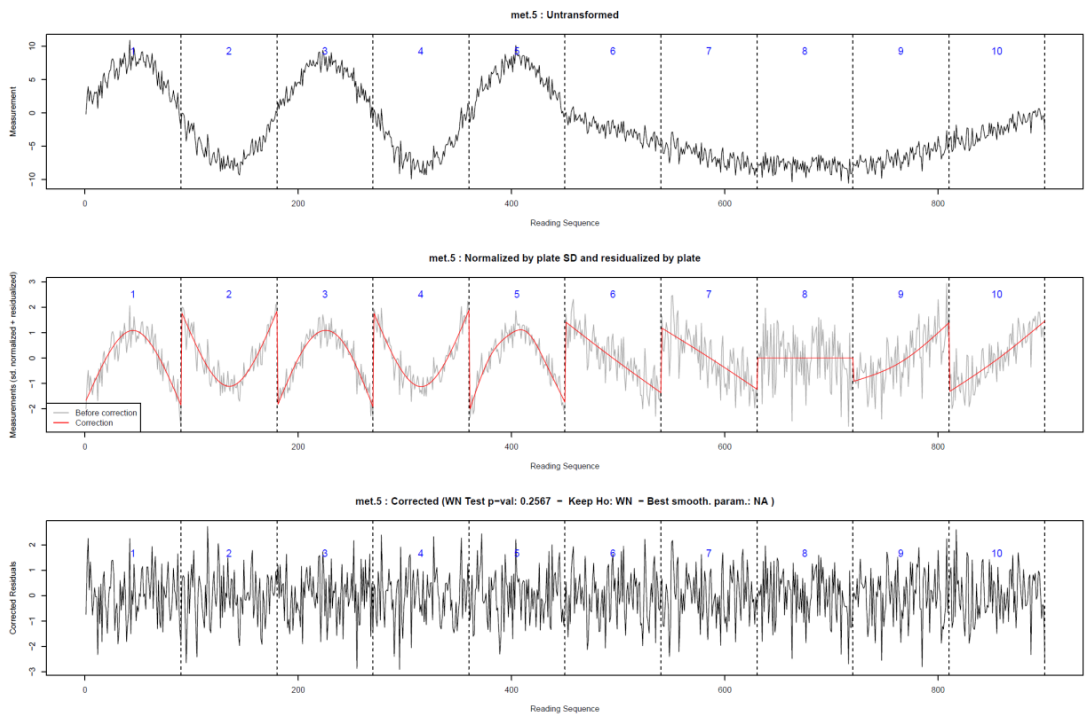

**Simulated true signal as N(0,1) random variable, added batch effects and/or drifts to it. Top panels: true signal distorted by batch effects and drifts; middle panels – intermediate WiNNbeta steps: WiNNbeta filtered out batch effects (black) and estimated drift correction (red); bottom panels: signal after batch effects and drifts were removed by WiNNbeta.**

April 11, 2024

\

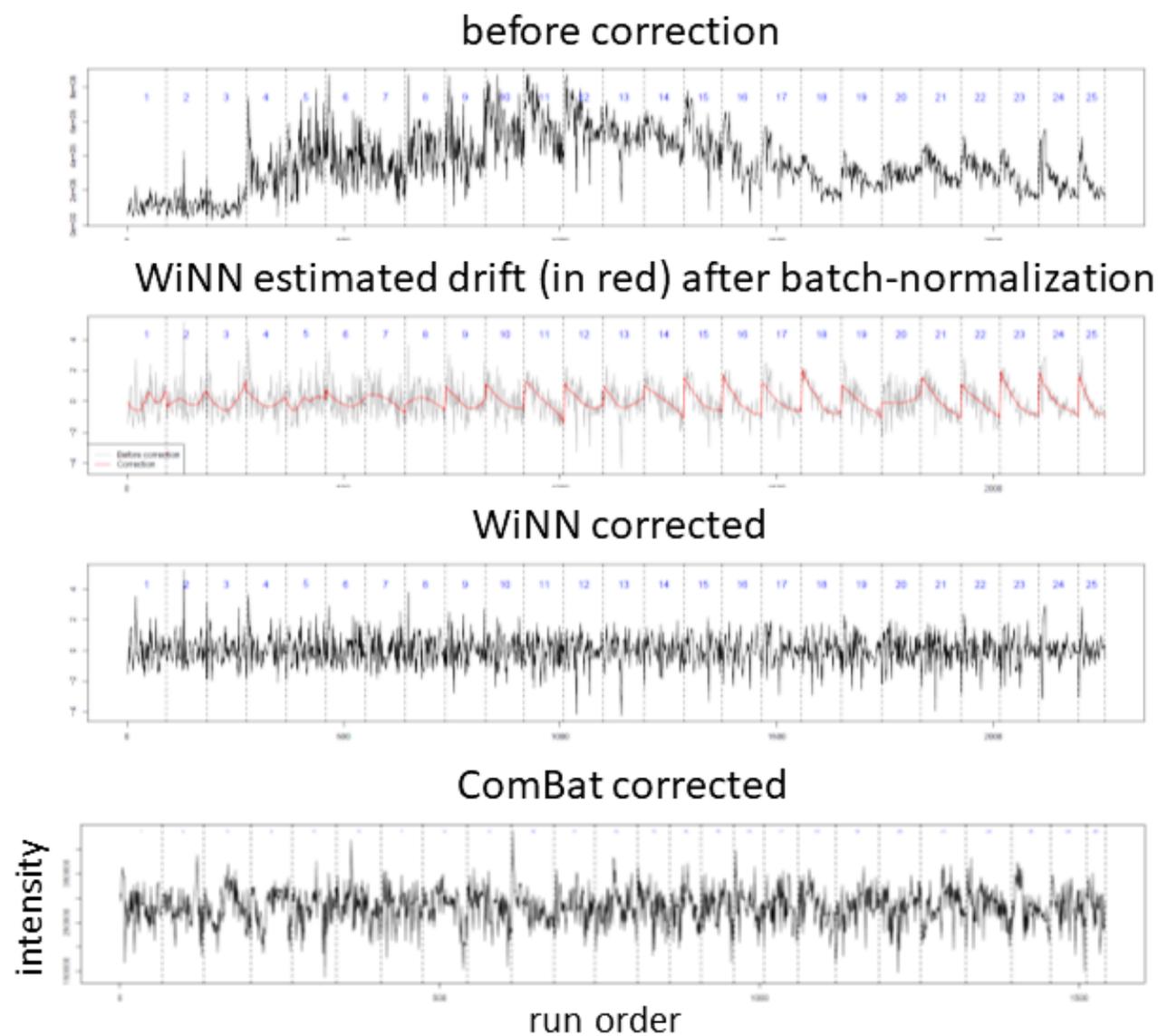

**Figure A2. WiNNbeta versus ComBat correction of one LC-MS metabolite**

April 11, 2024

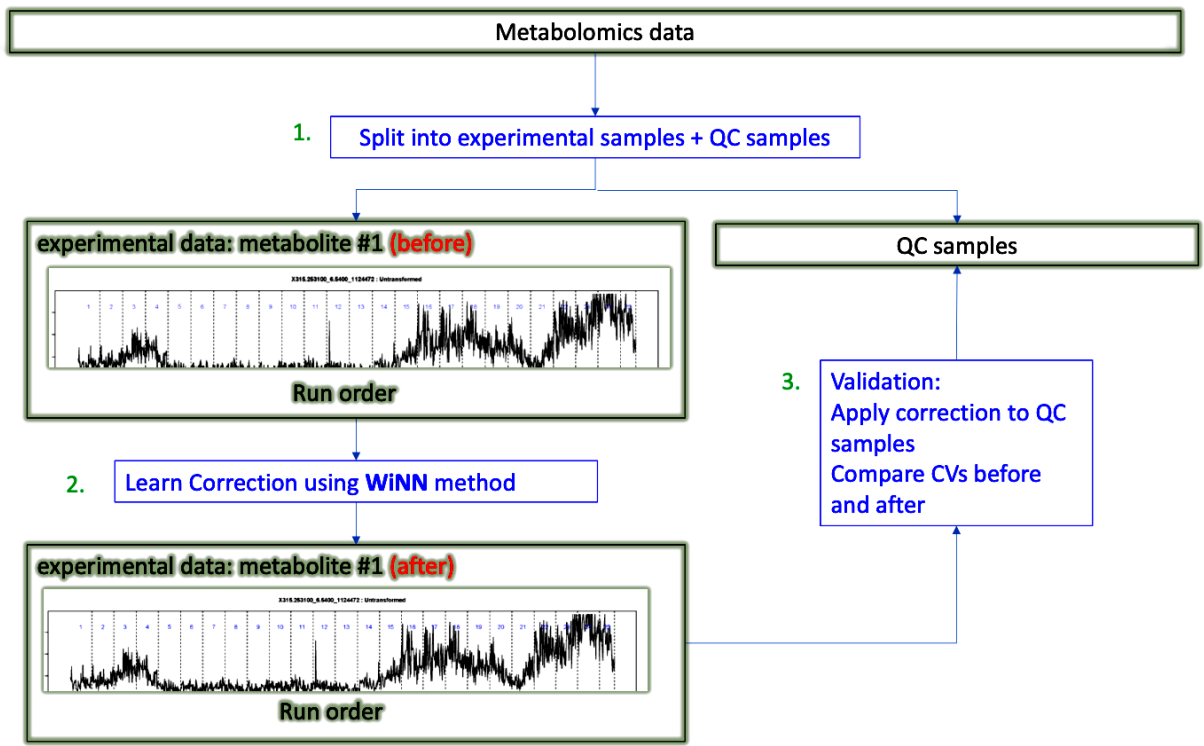

**Figure A3. Outline of how to use calculate CV before and after WiNNbeta correction.** Quality control samples are set aside and WiNNbeta method is applied to experimental data only and to each metabolite in particular. Then correction calculated in experimental data was applied to QC samples. Correction of QC samples consists of performing same shifting and rescaling operations informed by WiNNbeta in experimental data.

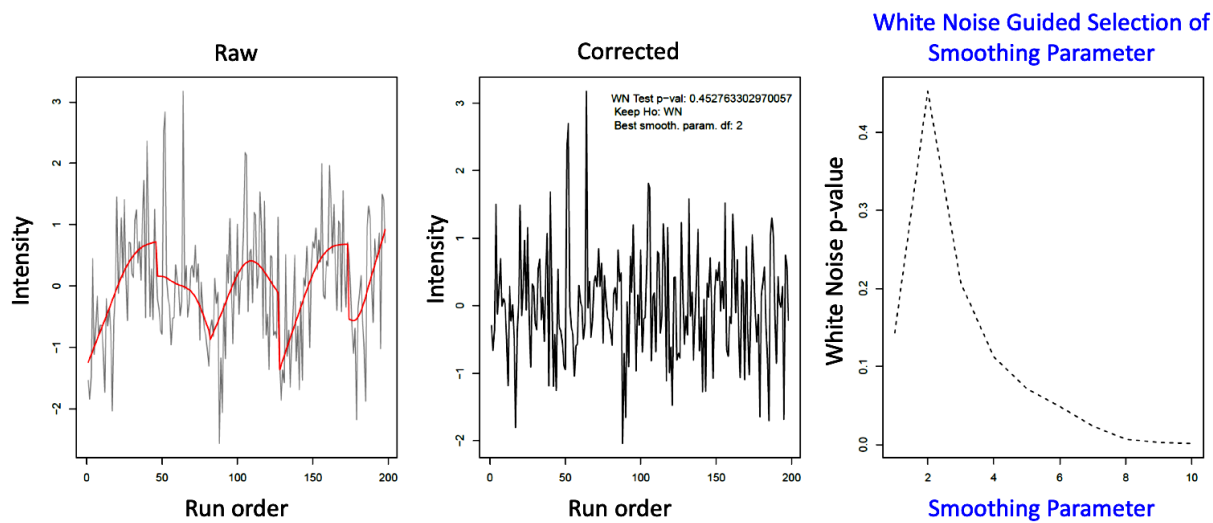

**Figure A4. Finding optimal smoothing parameters for drift-correction using the iid-guided smoothing splines.**

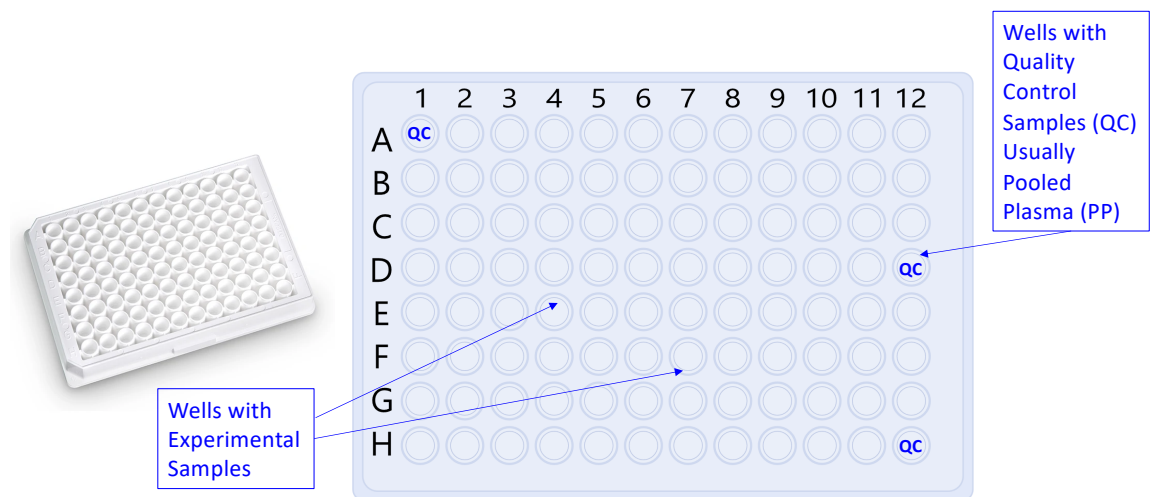

**Figure A5. Example of a 96-well plate and layout of experimental and quality control (QC) samples on a 96-well plate.** In a typical metabolomics experiment samples are aliquoted onto 94 or 384 well plate and processed one by one. QC samples are different from experimental samples and are placed at regular intervals.

April 11, 2024

**Appendix B.**

**B1. Drift correction with white noise-guided splines.**
To find the best spline smoothing parameter (denoted as $df$ in the Step (3) in Table 2) we use a drift-correction method by white noise-guided splines. We select $df$ that optimizes our white noise test. Because a good drift-correction will result in the larger p-value of white noise test, we select $df$ that maximizes white noise p-value. We apply a range of values of smoothing parameter and calculate for each the value of white noise test statistics for the residuals. Smoothing parameter that maximizes white noise statistic is considered optimal. Tuning of $df$ parameter for one metabolite is illustrated in Figure A4.

**B2. Evaluation of improvement in coefficients of variation before and after batch- and drift-correction using QC Samples.**
QC samples are inserted on each plate to monitor the quality of the extraction during the metabolomics experiment and evaluate measure of noise in the measurement of individual metabolites before analysis. One such measure is coefficient of variation (CV). It is defined as a ratio of the standard deviation (SD) of a metabolite to its mean ($CV = \frac{SD}{mean}$). Usually, CVs less than 20% are considered acceptable, i.e. as an indication that metabolite is measured precisely enoug[14]. Because QC samples were not used by WiNNbeta, we can use them to evaluate before and after correction CVs. Two points follow from the definition of CV above: 1) CV is defined only for metabolite intensities that cannot take negative values; 2) shifting the metabolite by a positive constant will artificially reduce its CV. While original uncorrected metabolite intensities are positive, our WiNNbeta correction algorithm often performs shifting in step (1) which may result in violation of the non-negativity assumption. Another issue is that QC samples can have means which are different from the mean of a given metabolite in experimental samples. For example, pooled plasma (PP) obtained from individuals from nursing homes may be enriched with fatty acids and therefore will fluctuate around a higher mean than in another cohort. Below we show how WiNNbeta correction can be applied to QC samples to overcome these issues. The algorithm is validated if the correction causes an enrichment of smaller CVs. Figure 5 shows the distribution of CVs before (orange) and after (green) the correction.

April 11, 2024

Table A3. Transfer WiNNbeta correction to QC samples (such as pooled plasma (PP))

For each metabolite $m$:

(1) **[Rescale $QC_m$ to match the mean in uncorrected experimental sample while preserving the CV]**

$$QC_m \rightarrow QC'_m$$

(2) **[Move back $y_m^{WiNN}$ to its original mean and standard deviation]** Shift and rescale intensities $y_m^{WiNN}$ of WiNNbeta-corrected metabolite $m$ to match the mean and standard deviation observed for $y_m$:

$$z_m = \left(y_m^{WiNN} + \underline{y_m} - \underline{y_m^{WiNN}}\right) \frac{SD(y_m)}{SD(y_m^{WiNN})}$$

(3) **[Calculate Correction]**

$$R = \frac{z_m}{y_m}$$

(4) **[Apply correction to $QC'_m$ (a metabolite from PP) by interpolating correction $R$ for the two closest experimental wells]**

$$QC_m^{WiNN} = R \cdot QC'_m$$

(5) **[Use $QC_m^{WiNN}$ to calculate CV after]**

The rescaling in step 1 in Table A3, which shifts the mean to the experimental one while preserving its CV is done as follows:

Define

$y_{QC}$ metabolite measured in QC samples before correction

$y'_{QC}$ metabolite measured in QC samples after correction

$y_e$ metabolite measured in experimental samples before correction

$y'_e$ metabolite measured in experimental samples after correction

$\underline{X}$ mean value of $X$ for any random variable $X$.

$SD(X)$ standard deviation of $X$ for any random variable $X$.

We have:

$CV(y_{QC}) = \frac{SD(y_{QC})}{\underline{y_{QC}}}$ and $CV(y'_{QC}) = \frac{SD(y'_{QC})}{\underline{y'_{QC}}}$

It can be shown that the linear transformation $y'_{QC} = a \cdot y_{QC} + b$ that matches the mean of $y_e$ while preserving its CV is given by:

$$a = \frac{CV(y_{QC}) \, \underline{y_e}}{SD(y_{QC})}$$

$$b = \underline{y_e} - \frac{CV(y_{QC})\,\underline{y_e}}{SD(y_{QC})}\,\underline{y_{QC}} = \underline{y_e} - a \cdot \underline{y_{QC}}.$$

Indeed: $y'_{QC} = ay_{QC} + b = ay_{QC} + \underline{y_e} - a \cdot \underline{y_{QC}} = a(y_{QC} - \underline{y_{QC}}) + \underline{y_e} \Rightarrow \underline{y'_{QC}} = a0 + \underline{y_e} = \underline{y_e}$

$$SD(y'_{QC}) = aSD(y_{QC}) = \frac{CV(y_{QC})\underline{y_e}}{SD(y_{QC})}SD(y_{QC}) = CV(y_{QC})\,\underline{y_e} = \frac{SD(y_{QC})}{\underline{y_{QC}}}\underline{y_e}$$

Therefore $CV(y'_{QC}) = \frac{\frac{SD(y_{QC})}{\underline{y_{QC}}}\underline{y_e}}{\underline{y_e}} = \frac{SD(y_{QC})}{\underline{y_{QC}}} = CV(y_{QC})$ and $a$ and $b$ define transformation of QC samples that preserves their CVs.

### B3. Experimental samples data preparation prior to running WiNNbeta

These are the pre-processing steps applied to the experimental baseline data:
1. Remove missing readings for a given metabolite or impute missing data
2. Truncate outliers for example by using 3 sigma rule: if $|metab - mean(metab)| \geq 3SD(metab)$, set $metab$ to some constant value or remove outliers.

### References

1. Dunn WB, Broadhurst D, Begley P, et al. Procedures for large-scale metabolic profiling of serum and plasma using gas chromatography and liquid chromatography coupled to mass spectrometry. Nature protocols 2011;6:1060-83.
2. Wishart DS. Emerging applications of metabolomics in drug discovery and precision medicine. Nature reviews Drug discovery 2016;15:473.
3. Livera AMD, Sysi-Aho M, Jacob L, et al. Statistical methods for handling unwanted variation in metabolomics data. Analytical chemistry 2015;87:3606-15.
4. Watrous JD, Niiranen TJ, Lagerborg KA, et al. Directed non-targeted mass spectrometry and chemical networking for discovery of eicosanoids and related oxylipins. Cell chemical biology 2019;26:433-42. e4.
5. Watrous JD, Henglin M, Claggett B, et al. Visualization, quantification, and alignment of spectral drift in population scale untargeted metabolomics data. Analytical chemistry 2017;89:1399-404.
6. Reisetter AC, Muehlbauer MJ, Bain JR, et al. Mixture model normalization for non-targeted gas chromatography/mass spectrometry metabolomics data. BMC bioinformatics 2017;18:1-17.
7. Sysi-Aho M, Katajamaa M, Yetukuri L, Orešič M. Normalization method for metabolomics data using optimal selection of multiple internal standards. BMC bioinformatics 2007;8:1-17.
8. Karpievitch YV, Nikolic SB, Wilson R, Sharman JE, Edwards LM. Metabolomics data normalization with EigenMS. PloS one 2014;9:e116221.
9. Karpievitch YV, Taverner T, Adkins JN, et al. Normalization of peak intensities in bottom-up MS-based proteomics using singular value decomposition. Bioinformatics 2009;25:2573-80.
10. Liu Q, Walker D, Uppal K, et al. Addressing the batch effect issue for LC/MS metabolomics data in data preprocessing. Scientific reports 2020;10:1-13.
11. Fan S, Kind T, Cajka T, et al. Systematic error removal using random forest for normalizing large-scale untargeted lipidomics data. Analytical chemistry 2019;91:3590-6.
12. Team RC. R: A Language and Environment for Statistical Computing. Vienna, Austria 2020.
13. Townsend MK, Clish CB, Kraft P, et al. Reproducibility of metabolomic profiles among men and women in 2

large cohort studies. Clinical chemistry 2013;59:1657-67.
14.	Health U.Do, Services H. Bioanalytical method validation, guidance for industry. http://www fda gov/cder/guidance/4252fnl htm 2001.
15.	Hinton G, Roweis ST. Stochastic neighbor embedding.  NIPS; 2002: Citeseer. p. 833-40.
16.	Van der Maaten L, Hinton G. Visualizing data using t-SNE. Journal of machine learning research 2008;9.
17.	Johnson WE, Li C, Rabinovic A. Adjusting batch effects in microarray expression data using empirical Bayes methods. Biostatistics 2007;8:118-27.